\documentclass[12pt]{article}

\usepackage{epsfig}

\def\ltwid{\mathrel{\raise.3ex\hbox{$<$\kern-.75em\lower1ex\hbox{$\sim$}}}}

\def\square{\kern1pt\vbox{\hrule height 1.2pt\hbox{\vrule width 1.2pt\hskip 3pt
   \vbox{\vskip 6pt}\hskip 3pt\vrule width 0.6pt}\hrule height 0.6pt}\kern1pt}
\def\ba{\begin{eqnarray}}
\def\ea{\end{eqnarray}}
\def\zetaa{\zeta_{1}}
\def\zetaaa{\zeta_{2}}
\def\zetaaaa{\zeta_{3}}
\def\alphaa{\alpha_{1}}
\def\alphaaa{\alpha_{2}}

\begin{document}

\begin{titlepage}
\begin{flushright}
UFIFT-QG-08-07
\end{flushright}

\vspace{0.5cm}

\begin{center}
\bf{Reconstructing the Distortion Function for Nonlocal Cosmology}
\end{center}

\vspace{0.3cm}

\begin{center}
C. Deffayet$^{*}$
\end{center}
\begin{center}
\it{APC \\UMR 7164
(CNRS, Univ. Denis Diderot-Paris 7, CEA, Obs. de Paris)\\
10 rue Alice Domon et L\'eonie Duquet, 75205 Paris Cedex 13, FRANCE}
\end{center}

\vspace{0.2cm}

\begin{center}
R. P. Woodard$^{\dagger}$
\end{center}
\begin{center}
\it{Department of Physics, University of Florida \\
Gainesville, FL 32611, UNITED STATES.}
\end{center}

\vspace{0.3cm}

\begin{center}
ABSTRACT
\end{center}
\hspace{0.3cm} We consider the cosmology of modified gravity models
in which Newton's constant is distorted by a function of the inverse
d'Alembertian acting on the Ricci scalar. We derive a technique for
choosing the distortion function so as to fit an arbitrary expansion
history. This technique is applied numerically to the case of 
$\Lambda$CDM cosmology, and the result agrees well with a simple
hyperbolic tangent.

\vspace{0.3cm}

\begin{flushleft}
PACS numbers: 04.30.-m, 04.62.+v, 98.80.Cq
\end{flushleft}

\vspace{0.1cm}

\begin{flushleft}
$^{*}$ deffayet@iap.fr \\
$^{\dagger}$ woodard@phys.ufl.edu
\end{flushleft}

\vspace{1cm}

\noindent {\it Dedicated to Stanley Deser on the occasion of his 78th birthday.}

\end{titlepage}

\section{Introduction}

Evidences in favor of an accelerating cosmic expansion are now quite 
numerous, reaching from the first systematic observations of Type Ia 
supernovae \cite{SNIA} to the more recent WMAP survey of the Cosmic 
Microwave Background \cite{WMAP}. The standard interpretation of this 
acceleration, in the form of a cosmological constant \cite{YW}, raises 
major well known questions. In particular it is not understood why the 
observed value of the vacuum energy density associated with the cosmological 
constant is so close to the non relativistic matter energy density (the 
so called coincidence problem) and so far from the ``natural'' order of 
magnitude expected from high energy physics \cite{SMC}. 

The problem is that the geometry reconstructed from observation is not 
sourced by known matter according to the Einstein equation,
\begin{equation}
\Bigl( R_{\mu\nu} - \frac12 g_{\mu\nu} R\Bigr)_{\rm obs} \neq \Bigl(8\pi G_N
T_{\mu\nu}\Bigr)_{\rm known} \; . \label{ineqn}
\end{equation}
Different ways addressing this problem can be classified as to whether it
is the right hand (matter) side or the left hand (gravity) side of 
(\ref{ineqn}) which is modified. ``Dark energy'' models explain the data
by introducing a new source of stress energy to the right hand side of
(\ref{ineqn}). A cosmological constant is one example; another is a scalar
quintessence field $\varphi$ described by the Lagrangian \cite{phi},
\begin{equation}
\mathcal{L} = -\frac12 \partial_{\mu} \varphi \partial_{\nu} \varphi 
g^{\mu\nu} \sqrt{-g} - V(\varphi) \sqrt{-g} \; . \label{quint}
\end{equation}
Local, metric-based modifications to the left hand side of (\ref{ineqn})
are restricted by stability \cite{RPW} to take the form of replacing the
Ricci scalar of the Hilbert Lagrangian by an arbitrary function of the
Ricci scalar \cite{f(R),CNO},
\begin{equation}
-\frac1{16 \pi G_N} \, R \sqrt{-g} \longrightarrow -\frac1{16 \pi G_N} \, 
F(R) \sqrt{g} \; . \label{FGR}
\end{equation}
More general modifications of gravity must either abandon stability or locality
\cite{TW,nonlocal,SW,DW}, or they must involve some field other than the 
metric to carry part of the gravitational force \cite{TeVeS,STV}. Other 
proposals involving extra dimensions or massive gravitons have not yet 
reached a fully satisfactory state \cite{DGP,DEGRAV}.

Central to the evaluation of any class of models is the {\it reconstruction
problem}. This consists of identifying the extent to which parameters in the
model such as $V(\varphi)$ and $F(R)$ can be adjusted to support a geometry 
of the Friedmann-Lema\^{\i}tre-Robertson-Walker (FLRW) form with flat spatial 
sections,
\begin{equation}
ds^2 = -dt^2 + a^2(t) d\vec{x} \! \cdot \! d\vec{x} \; , \label{FLRW}
\end{equation}
where the scale factor $a(t)$ is a known but arbitrary function of time.
Of course there is just one expansion history in nature and a putative 
model need only explain that, but if reconstruction can be solved for
a general scale factor then it is certain a model within the given class 
can fit the actual expansion history. To the extent that the solution is
constructive one also obtains important constraints on the model, of 
course limited by the precision with which the expansion history can be
measured.

It is straightforward to solve the reconstruction problem for scalar 
quint\-es\-sence models \cite{TW,SRSS,CNO} and a brief presentation of the
solution will help focus ideas. For the FLRW geometry (\ref{FLRW}) the 
scalar must be independent of space, and only two of Einstein's equations 
are nontrivial,
\begin{eqnarray}
3 H^2 & = & 8\pi G_N \Bigl[ \frac12 \dot{\varphi}^2 + V(\varphi)\Bigr] \; ,
\label{E1} \\
-2 \dot{H} - 3 H^2 & = & 8\pi G_N \Bigl[ \frac12 \dot{\varphi}^2 - 
V(\varphi)\Bigr] \; . \label{E2}
\end{eqnarray}
Here and henceforth, $G_N$ is the Newton constant, a dot means a derivative 
with respect to the cosmic time $t$ and we shall henceforth employ the usual 
definition of the Hubble parameter, $H \equiv \dot{a}/a$. By adding 
(\ref{E1}) and (\ref{E2}) one obtains the relation,
\begin{equation} \nonumber
-2\dot{H} = 8 \pi G_N \, \dot{\varphi}^2 \; .
\end{equation}
Hence one can reconstruct the scalar's evolution, and even invert it to
express time in terms of the scalar, provided the Hubble parameter is 
monotonically decreasing, 
\begin{equation}
\varphi(t) = \varphi_0 \pm \int_0^{t} \!\! du \sqrt{\frac{-2 \dot{H}(u)}{
8\pi G_N}} \qquad \Longleftrightarrow \qquad t = t(\varphi) \; .
\end{equation}
One then determines the potential by subtracting (\ref{E2}) from (\ref{E1})
and evaluating the resulting (assumed known) function of time at $t(\varphi)$,
\begin{equation}
\Bigl[2 \dot{H}(t) + 6 H^2(t) \Bigr]_{t(\varphi)} = 16 \pi G_N \, V(\varphi)
\; .
\end{equation}

Similar reconstruction procedures exist for $F(R)$ models \cite{RPW,CNO}.
The purpose of this paper is to solve the reconstruction problem for a
recently proposed nonlocal cosmology model in which one multiplies
the Hilbert Lagrangian by an algebraic function of the inverse scalar
d'Alembertian acting on the Ricci scalar \cite{DW},
\begin{equation}
{\cal L}_g \equiv \frac1{16 \pi G_N} \, \sqrt{-g} R \left[ 1+ 
f\Bigl(\frac1{\square} R\Bigr) \right]\; , \label{DL2}
\end{equation}
The motivation for this class of models is to trigger late time 
acceleration by the transition from radiation domination, during which
the Ricci scalar is nearly zero, to matter domination at about $10^5$
years after the Big Bang. The subsequent time lag to the observed onset
of acceleration, at about $10^9$ years, would be provided by the effect 
of the transition being reflected through the nonlocal, inverse 
d'Alembertian. Although there is hope of eventually deriving a model of 
the class (\ref{DL2}) from quantum field theoretic loop corrections, the 
proposal is at this stage purely phenomenological.

The free parameter $f(X)$ in expression (\ref{DL2}) is known as the 
{\it nonlocal distortion function}. Absent a derivation from fundamental
theory, it has the same status as the potential $V(\varphi)$ in (\ref{quint}) 
and the function $F(R)$ in (\ref{FGR}). What we will do in this paper is
first to show how $f(X)$ can be tuned to give an arbitrary $a(t)$, then we
will work out the specific form $f(X)$ must take to reproduce the $a(t)$ of
$\Lambda$CDM, without actually employing a cosmological constant. This 
problem has already been studied in an excellent paper by Koivisto \cite{TK1} 
(see also \cite{TK2}) but for a local variant of the model, introduced by 
Nojiri and Odintsov \cite{NO1}, in which a scalar Lagrange multiplier forces 
the d'Alembertian of a second scalar give $R$. That version of the model has 
additional degrees of freedom that the original proposal lacks \cite{NAK}, 
so it is important to examine the reconstruction problem in both formulations.

This paper is organized as follows. In section \ref{sec2} we summarize the 
model of Ref. \cite{DW} and its specialization to cosmology. In section 
\ref{sec3} we outline the steps of the reconstruction of the function $f(X)$ 
once a given cosmology is chosen. In section \ref{sec4}, we show how to 
obtain the function $f$ suitable to reproduce the $\Lambda$CDM cosmological 
evolution with the same matter content but no cosmological constant. We then
summarize our results and conclude in section \ref{sec5}.

\section{Nonlocal Cosmology}
\label{sec2}

The modified gravity Lagrangian (\ref{DL2}) has already been given.
It remains to state that $\square$ is the covariant scalar d'Alembertian,
\begin{equation} \nonumber
\square \equiv \frac1{\sqrt{-g}} \, \partial_{\mu} \Bigl(\sqrt{-g}
g^{\mu\nu} \partial_{\nu} \Bigr) \; .
\end{equation}
By its inverse we mean the retarded Green's function. For the FLRW 
geometry (\ref{FLRW}) the action of this inverse on some function of time 
$W(t)$ takes the simple form,
\begin{equation} \label{1BH}
\frac1{\square} \Bigl[W\Bigr](t) = -\int_0^t \!\! du \, \frac1{a^3(u)}
\int_0^{u} \!\! dv \, a^3(v) W(v) \; .
\end{equation}
The metric $g_{\mu \nu}$ is assumed to be minimally coupled to matter. 

The model is actually defined by its field equations, which are obtained
by varying the gravity and matter actions with respect to the metric and
then employing the partial integration trick explained in \cite{SW}. This
produces causal and conserved field equations, like those of the more rigorous
Schwinger-Keldysh formalism \cite{RJ}. These equations take the form,
\ba \label{EQ1}
G_{\mu \nu} + \Delta G_{\mu \nu} = 8 \pi G_N T_{\mu \nu},
\ea
where $T_{\mu \nu}$ is the matter total  energy momentum tensor (including 
a possibly non vanishing cosmological constant), and $\Delta G_{\mu \nu}$ 
comes from varying the non local $f$ term in the above action (\ref{DL2}). 

In the following, we will restrict the metric to be of FLRW form (\ref{FLRW}).
With this ansatz, the field equations (\ref{EQ1}) take the form 
\ba
3H^2  + \Delta G_{00} &=& 8 \pi G \rho \label{EQrho} , \\
-2 \dot{H} - 3H^2  + \frac{1}{3 a^2} \delta^{ij} \Delta G_{ij} 
&=& 8 \pi G P , \label{EQP}
\ea
$\rho$ and $P$ being respectively the matter total energy density and pressure, and the non local pieces reading
\begin{eqnarray}
\lefteqn{\Delta G_{00} = \Bigl[3 H^2 + 3 H \partial_{t}\Bigr] 
\Biggl\{ f \Bigl(\frac1{\square} R\Bigr) + \frac1{\square} 
\Biggl[ R f'\Bigl(\frac1{\square} R\Bigr)\Biggr] \Bigg\} } \nonumber \\
& & \hspace{5cm} + \frac12 \partial_{t} \Bigl(\frac1{\square} R\Bigr) \times 
\partial_{t} \Biggl(\frac1{\square} \Biggl[ R f'\Bigl(\frac1{\square} R 
\Bigr) \Biggr] \Biggr) , \qquad \label{D200} \\
\lefteqn{\Delta G_{ij} = -\Bigl[2 \dot{H} + 3 H^2 + 2 H \partial_{t} + 
\partial_{t}^2\Bigr] \Biggl\{ f\Bigl(\frac1{\square} R\Bigr) + 
\frac1{\square} \Biggl[ R f'\Bigl(\frac1{\square} R\Bigr)\Biggr] \Bigg\} 
g_{ij} } \nonumber \\
& & \hspace{5cm} + \frac12 \partial_{t} \Bigl(\frac1{\square} R\Bigr) \times 
\partial_{t} \Biggl(\frac1{\square} \Biggl[ R f'\Bigl(\frac1{\square} R 
\Bigr) \Biggr] \Biggr) g_{ij} \; , \qquad \label{D2ij}
\end{eqnarray}
where $f'$ denotes the derivative of $f$ with respect to its argument.
As already stressed, the left hand side of equation (\ref{EQ1}) is 
conserved, and hence the first Friedmann equation (\ref{EQrho}) and the 
matter energy-momentum conservation equation, 
\ba \label{CONSERV}
\dot{\rho} + 3H(P + \rho)=0,
\ea 
are enough to ensure that equation (\ref{EQP}) is fulfilled, as is the 
case with standard Friedmann equations. 

\section{General Reconstruction Technique}
\label{sec3}

We first note that the difference between field equations (\ref{EQrho}) 
and (\ref{EQP}) leads to a simple second order ODE for the function $F$ 
defined as 
\ba \label{Ff2}
F &=& f+\frac{1}{\square}\left(R \frac{d f}{dX}
\right), 
\ea
where $X$ is defined as
\ba \label{DEFX}
X \equiv \frac{1}{\square}{R}.
\ea
This ODE reads 
\ba \label{ODE}
\ddot{F} +5 H\dot{F} + \left(6 H^2 + 2 \dot{H}\right) \left(F+1\right) = 8 \pi G_N\left(\rho-P\right).
\ea
If one then assumes the matter content of the Universe (specified here-above 
by its total energy density $\rho$ and pressure $P$) and its cosmological 
evolution (specified by some scale factor $a(t)$) to be chosen, the first 
step to reconstruct $f$ is to solve Eq. (\ref{ODE}), which allows to obtain 
$F$ as a function of the cosmological time $t$. Then we invert relation 
(\ref{Ff2}) rewritten as 
\ba \label{SQUAREF} 
\square F = \square f + R \dot{f} \frac{dX}{dt}, 
\ea
and yielding thus an ODE provided $X$ is known as a function of $t$ via 
equation (\ref{DEFX}). This allows to obtain $f$ as a function\footnote{
Denoting here by the same letter the function $f(t)\equiv f(X(t))$ and the 
function $f(X)$} of $t$. The last step is to invert $X(t)$ to obtain $t$ 
as a function of $X$. This allows to get $f(X)$, the function which appears 
in the original action (\ref{DL2}). 

In this process, some care has to be taken about the boundary conditions. 
First, we emphasize that our choice of the retarded Green function in the 
definition of the inverse of the d'Alembertian operator (\ref{1BH}) does 
not permit inclusion of the extra zero modes that cause the ambiguities 
underlined in Ref. \cite{NAK} about the local version of the model 
considered in \cite{TK1,NO1}. Moreover, one should make sure that whatever 
boundary conditions get imposed to integrate the ODEs are compatible with 
both equations (\ref{EQrho}) and (\ref{EQP}). The need for this might seem 
confusing in view of the close relation between the two equations implied 
by conservation,
\begin{equation}
\Bigl[ \frac{d}{dt} + 3 H\Bigr] \Bigl({\rm Eqn.}~\ref{EQrho}\Bigr) = 
-3 H \times \Bigl({\rm Eqn.}~\ref{EQP}\Bigr) \; . \label{cons}
\end{equation}
However, it will be noted that relation (\ref{cons}) involves a derivative 
of equation (\ref{EQrho}). Had we based the reconstruction technique solely
upon equation (\ref{EQrho}) then equation (\ref{EQP}) would have followed
automatically. But our reconstruction procedure instead employs the
difference of equations (\ref{EQrho}) and (\ref{EQP}), and this difference
only defines equation (\ref{EQrho}) up to an integration constant.

In the next section, we carry out these steps for the special case where 
the matter content of the Universe only consists of non relativistic and 
relativistic matter without any cosmological constant, while the 
cosmological evolution is that given by usual Friedmann equation with the 
same matter content plus a non vanishing cosmological constant, i.e. the 
one of the $\Lambda$CDM model. 

\section{Specialization to $\Lambda$CDM}
\label{sec4}

We want to reproduce $\Lambda$CDM cosmology with the same matter content 
but a vanishing cosmological constant. Hence we assume that the Hubble 
parameter $H$, appearing in equation (\ref{ODE}), is a solution of the 
standard Friedmann equations {\it with} a cosmological constant and the 
same matter content as in our Universe: 
\ba
3H^2 - \Lambda &=& 8 \pi G \rho, \label{Frie1} \\
2 \dot{H} + 3H^2 - \Lambda  &=& -8 \pi G P. \label{Frie2}
\ea
It is then easy to see that Eq
(\ref{ODE}) simplifies dramatically to read now 
\begin{equation}
\ddot{F} +5 H\dot{F} + \left(6 H^2 + 2 \dot{H}\right)F = 
- 6 H_0^2 \Omega_\Lambda , \label{Fzeqnbis}
\end{equation}
where  $H_0$ is the Hubble parameter today, and $\Omega_\Lambda$ is 
defined as usual in terms of the cosmological constant $\Lambda$ entering 
into equations (\ref{Frie1}-\ref{Frie2}), i.e. as 
\ba
\Omega_\Lambda = \frac{\Lambda}{3 H_0^2}.
\ea
Notice that the matter energy density and pressure appearing on the 
right hand side of equation (\ref{ODE}) have cancelled against 
$6 H^2 + 2 \dot{H}$ appearing on the left hand side.

In the rest of the paper, we will further simplify the problem by 
considering that the only matter content of the Universe is a sum of 
two components, one of non relativistic matter (with $\Omega$ parameter 
$\Omega_m$) and one of relativistic matter (with $\Omega$ parameter 
$\Omega_r$). It will then turn out to be convenient to use equations 
(\ref{Frie1}) and (\ref{Frie2}) to reexpress $H$ and its time derivatives 
in terms of $\Omega_\Lambda$,  $\Omega_r$ and $\Omega_m$ and to use 
the variable $\zeta$ instead of $t$, defined in terms of the redshift $z$ as 
\begin{equation} \nonumber
\zeta \equiv 1 + z = \frac1{a(t)}.
\end{equation}
We also introduce the dimensionless Hubble parameter $h(\zeta)$ given by 
\ba \label{hdef} 
h(\zeta) = \sqrt{\Omega_{\Lambda} + \Omega_m \zeta^3 + \Omega_r \zeta^4}, 
= H/H_0.
\ea

Equation (\ref{Fzeqnbis}) can be readily integrated\footnote{Using that 
the fact that $\zeta^2$ is a homogeneous solution.} to yield $F$ in the form 
\ba \label{FPHI}
F\equiv \zeta^2 \Phi^2,
\ea
where $\Phi$ is given by 
\begin{eqnarray}
\Phi(\zeta) &=& \Phi_{\rm eq} + h(\zeta_{\rm eq}) \Phi^\prime_{\rm eq} 
\int_{\zeta_{\rm eq}}^\zeta \frac{d \zetaa}{h(\zetaa)}\nonumber  
-6 \Omega_\Lambda \int_{\zeta_{\rm eq}}^\zeta \frac{d \zetaa}{h(\zetaa)} 
\int_{\zeta_{\rm eq}}^{\zetaa} \frac{d \zetaaa}{\zetaaa^{4} h(\zetaaa)},
\end{eqnarray}
and the integration constants $\Phi_{\rm eq}$ and  $\Phi^\prime_{\rm eq}$ 
are defined respectively as the values of $\Phi$ and $\Phi^\prime$ (here 
and in the following, a prime denotes a derivative with respect to $\zeta$) 
at the matter-radiation equality $\zeta_{\rm eq}$. In order to have a well 
behaved $F$ at early times, we demand that $F(\zeta)$ goes to zero at large 
$\zeta$.  This fixes $\Phi_{\rm eq}$ and  $\Phi^\prime_{\rm eq}$ to be 
\ba \label{choice1}
\Phi_{\rm eq} &=& -6\Omega_{\Lambda} \int_{\zeta_{\rm eq}}^{\infty} \!\! d\zetaa \,
\frac1{h(\zetaa)} \int_{\zetaa}^{\infty} \!\! d\zetaaa \, 
\frac1{\zetaaa^{4} h(\zetaaa)} \;.\\ \label{choice2}
\Phi_{\rm eq}' &=& \frac{6\Omega_{\Lambda}}{h_{\rm eq}}  \int_{\zeta_{\rm eq}}^{\infty} \!\! d\zetaa \, 
\frac1{\zetaa^{  4} h(\zetaa)}.
\ea
The resulting expression for $\Phi(\zeta)$ is then given by 
\begin{equation} \nonumber
\Phi(\zeta) = -6\Omega_{\Lambda} \int_{\zeta}^{\infty} \!\! d\zetaa \,
\frac1{h(\zetaa)} \int_{\zetaa}^{\infty} \!\! d\zetaaa \, 
\frac1{\zetaaa^{4} h(\zetaaa)} \;,
\end{equation}
and the large $\zeta$ expansion of $\Phi(\zeta)$ is,
\begin{equation} \nonumber
\Phi(\zeta) = -\frac{\Omega_{\Lambda}}{5 \Omega_r} \, \frac1{\zeta^6} +
{\cal O}\Bigl(\frac1{\zeta^7}\Bigr) \; .
\end{equation}
Any choice other than (\ref{choice1}-\ref{choice2}) will result in a function 
$F(\zeta)$ actually growing for large $\zeta$.

Having obtained $F(\zeta)$ we now turn to invert equation (\ref{Ff2}).  
This equation reads, 
\ba 
\frac{\zeta^2 h^2}{(2 \Omega_\Lambda + \frac{1}{2} \Omega_m \zeta^3)} 
\frac{d^2}{d \zeta^2}\left(f-F\right) - \frac{df}{d\zeta}
\left(\zeta + 6 \frac{d \zeta}{dX}\right) = - \zeta \frac{dF}{d\zeta}. 
\ea
Knowing $F$ from (\ref{FPHI}) it can easily be integrated once to yield 
the general solution 
\begin{eqnarray}
\lefteqn{ \frac{df}{d\zeta} = 2 \zeta \Phi(\zeta) + \frac{\zeta^2}{h(\zeta)
I(\zeta)} \Biggl\{ \Bigl[ (f')_{eq} \!-\! 2 \Phi_{eq}\Bigr] \frac{h_{eq} 
I_{eq}}{\zeta_{eq}^2} } \nonumber \\
& & \hspace{1.5cm} + 6 \Omega_{\Lambda} \int_{\zeta}^{\zeta_{eq}} \!\! d\zetaa 
\, \frac{I(\zetaa)}{\zetaa^{4} h(\zetaa)} - 2 \int_{\zeta}^{\zeta_{eq}}
\!\! d\zetaa \, \frac{(12 \Omega_{\Lambda} \!+\! 3 \Omega_{m}\zetaa^{3})
\Phi(\zetaa)}{\zetaa^{5}} \Biggr\} \;.\qquad
\end{eqnarray}
This expression contains a new integration constant, $(f')_{eq}$, and the 
function $I(\zeta)$ defined by 
\ba \label{Idef}
I(\zeta) \equiv \int^\infty_\zeta \frac{d\zetaa}{\zetaa^{4}H_0 H(\zetaa)} R,
= \int_{\zeta}^{\infty} \!\! d\zetaa \, \frac{(12 
\Omega_{\Lambda} \!+\! 3 \Omega_{m} \zetaa^{3})}{\zetaa^{4}
h(\zetaa)} \; .
\ea
Now, one also has from equation (\ref{Ff2}) that 
\begin{eqnarray}
F(\zeta) & = & f\Bigl(X(\zeta)\Bigr)  \nonumber
-\int_{\zeta}^{\infty} \!\! d\zetaa \, \frac{\zetaa^{2}}{h(\zetaa)} \int_{\zetaa}^{\infty} \!\! d\zetaaa \, \frac{(12 \Omega_{\Lambda} 
\!+\! 3 \Omega_{m} \zetaaa^{3})}{\zetaaa^{4} h(\zetaaa)} 
\times \frac{d f}{d X} \; , \qquad \\
& = & f\Bigl(X(\zeta)\Bigr) 
-\int_{\zeta}^{\infty} \!\! d\zetaa \, \frac{\zetaa^{2}}{h(\zetaa)} 
\int_{\zetaa}^{\infty} \!\! d\zetaaa \, \frac{(12 \Omega_{\Lambda} \!+\! 
3 \Omega_{m} \zetaaa^{3})}{\zetaaa^{6} I(\zetaaa)} 
\times \frac{d f}{d \zetaaa} \; . \qquad \label{def}
\end{eqnarray}
where we have used the 
the definition of $\square^{-1}$ given in equation (\ref{1BH}). In particular, we have that 
\begin{equation} \nonumber
 \frac1{\square} \Bigl[ R\Bigr] =
-\int_{\zeta}^{\infty} \!\! d\zetaa \, \frac{\zetaa^{2}}{
h(\zetaa)} \int_{\zetaa}^{\infty} \!\! d\zetaaa \, 
\frac{(12 \Omega_{\Lambda} \!+\! 3 \Omega_{m} \zetaaa^{3})}{
\zetaaa^{4} h(\zetaaa)} \;= X(\zeta).
\end{equation}
which was used to obtain equation (\ref{def}).  
The integral in Eq. (\ref{def}) will only make sense provided 
$df/d\zeta$ falls off faster that $1/\zeta^2$ for large $\zeta$.
This fixes the new integration constant $(f')_{\rm eq}$ to be 
\begin{eqnarray}
\lefteqn{(f')_{eq} = 2 \Phi_{eq} } \nonumber \\
& & \hspace{.8cm}+\frac{\zeta_{eq}^2}{h_{eq} I_{eq}} \Biggl\{6 \Omega_{\Lambda}
\!\! \int_{\zeta_{eq}}^{\infty} \!\! d\zetaa \, \frac{I\zetaa)}{\zetaa^{4} h(\zetaa)} - 2 \!\! \int_{\zeta_{eq}}^{\infty} \!\! d\zetaa \, 
\frac{(12 \Omega_{\Lambda} \!+\! 3 \Omega_{m} \zetaa^{3}) \Phi(\zetaa)}{
\zetaa^{5}} \Biggr\} \; , \qquad
\end{eqnarray}
and results in the new expression for $df/d\zeta$ reading 
\begin{eqnarray}
\lefteqn{ \frac{df}{d\zeta} = 2 \zeta \Phi(\zeta) + \frac{\zeta^2}{h(\zeta)
I(\zeta)} \Biggl\{ 6 \Omega_{\Lambda} \int_{\zeta}^{\infty} \!\! d\zetaa 
\, \frac{I(\zetaa)}{\zetaa^{4} h(\zetaa)} } \nonumber \\
& & \hspace{5.5cm} - 2 \int_{\zeta}^{\infty} \!\! d\zetaa \, 
\frac{(12 \Omega_{\Lambda} \!+\! 3 \Omega_{m} \zetaa^{3})
\Phi(\zetaa)}{\zetaa^{5}} \Biggr\} \; . \qquad \label{f2prime}
\end{eqnarray}
If one integrates once more, one obtains another integration constant, 
namely the value of $f(\zeta)$ at $\zeta=\zeta_{\rm eq}$. This constant 
can be fixed demanding that relation (\ref{def}) hold at $\zeta = \zeta_{eq}$. 
We get
\begin{equation} \nonumber
(f)_{eq} = \zeta_{eq}^2 \Phi_{eq} - \int_{\zeta_{eq}}^{\infty} \!\! d\zetaa 
\, \frac{\zetaa^{2}}{h(\zetaa)} \int_{\zetaa}^{\infty} \!\! d\zetaaa \,
\frac{(12 \Omega_{\Lambda} \!+\! 3 \Omega_{m} \zetaaa^{3})}{
\zetaaa^{6} I(\zetaaa)} \times \frac{d f}{d \zetaaa} \; .
\end{equation}
Note that this relation is not self-referential for $(f)_{eq}$ because 
the factor of $d f/d\zeta''$ on the right hand side does not involve
$(f)_{eq}$. With such a choice for $(f)_{eq}$, the function $f\Bigl(X(\zeta)\Bigr)$ vanishes 
at early times, that is, for large $\zeta$. We can therefore construct $f$
by simply integrating equation (\ref{f2prime}) to obtain
\begin{eqnarray} \nonumber
\lefteqn{f \Bigl(X(\zeta)\Bigr) = -\int_{\zeta}^{\infty} \!\! d\zetaa \, 
\frac{d f}{d\zetaa} \; , } \\
& & = -2\int_{\zeta}^{\infty} \!\! d\zetaa \,\zetaa \Phi(\zetaa) 
- 6 \Omega_{\Lambda} \int_{\zeta}^{\infty} \!\! d\zetaa \, \frac{\zetaa^{
2}}{h(\zetaa) I(\zetaa)} \int_{\zetaa}^{\infty} \!\! d\zetaaa \,
\frac{I(\zetaaa)}{\zetaaa^{ 4} h(\zetaaa)} \qquad \nonumber \\ \nonumber
& & \hspace{1cm} + 2 \int_{\zeta}^{\infty} \!\! d\zetaa \, \frac{\zetaa^{
2}}{h(\zetaa) I(\zetaa)} \int_{\zetaa}^{\infty} \!\! d\zetaaa \,
\frac{(12 \Omega_{\Lambda} \!+\! 3 \Omega_{m} \zetaaa^{ 3}) 
\Phi(\zetaaa)}{\zetaaa^{5}} \; , \qquad \\
& & = 6\Omega_{\Lambda} \int_{\zeta}^{\infty} \!\! d\zetaa \, 
\frac{(\zetaa^{2} \!-\! \zeta^2)}{h(\zetaa)} \int_{\zetaa}^{\infty} \!\!
d\zetaaa \, \frac1{\zetaaa^{4} h(\zetaa)} \nonumber \\
& & \hspace{1cm} -6\Omega_{\Lambda} \int_{\zeta}^{\infty} \!\! d\zetaa \,
\frac{\zetaa^{2}}{h(\zetaa) I(\zetaa)} \int_{\zetaa}^{\infty} \!\!
d\zetaaa \, \frac{I(\zetaaa)}{\zetaaa^{ 4} h(\zetaaa)} \nonumber \\
& & \hspace{2cm} - 36 \Omega_{\Lambda} \int_{\zeta}^{\infty} \!\! d\zetaa \,
\frac{\zetaa^{2}}{h(\zetaa) I(\zetaa)} \int_{\zetaa}^{\infty} \!\!
d\zetaaa \, \Biggl[ \frac{(\Omega_{\Lambda} \!+\! \Omega_{m} 
\zetaa^{3})}{\zetaa^{4}} \nonumber \\
& & \hspace{4.5cm} - \frac{(\Omega_{\Lambda} \!+\! 
\Omega_{m} \zetaaa^{3})}{\zetaaa^{4}} \Biggr] 
\frac1{h(\zetaaa)} \int_{\zetaaa}^{\infty} \!\! d\zetaaaa \,
\frac1{\zetaaaa^{4} h(\zetaaaa)} \; . \qquad \label{ffin}
\end{eqnarray}
Introducing the variable $\alpha = \zeta_{eq}/\zeta$, as well as the 
parameter $\omega$ given by $\omega \equiv \Omega_{\Lambda} \Omega_{r}^3/
\Omega_{m}^4$, the above expression for $f$ can be given in term of the 
two elliptic integrals $\overline{J}(\alpha)$ and $\overline{I}(\alpha)$ 
defined by 
\begin{eqnarray}
\overline{J}(\alpha) & \equiv & \int_0^{\alpha} \!\! d\alphaa \, 
\frac{\alphaa^{4}}{\sqrt{1 \!+\! \alphaa \!+\! \omega \alphaa^{4}}}
\; , \label{Jbar} \\ 
\overline{I}(\alpha) & \equiv & \int_0^{\alpha} \!\! d\alphaa \, 
\frac{\alphaa \!+\! 4 \omega \alphaa^{4}}{\sqrt{1 \!+\! \alphaa \!+\! 
\omega \alphaa^{4}}} \label{Ibar} = \frac{\zeta_{eq}^{3/2}}{3 \Omega_m^{1/2}} 
I(\zeta). 
\end{eqnarray}
If we then define $f_1$, $f_2$ and $f_3$ as the three contribution to 
$f$ appearing in the right hand side of equation (\ref{ffin}), such that 
$f(\zeta)=f_1(\zeta)+f_2(\zeta)+f_3(\zeta)$, one has  
\begin{eqnarray} \label{f11}
f_1\Bigl(X(\zeta)\Bigr) 
& = & 6 \omega \int_0^{\alpha} \!\! d\alphaa \, 
\frac{(\frac1{\alphaa^{2}} \!-\! \frac1{\alpha^2}) \overline{J}(\alphaa)
}{\sqrt{1 \!+\! \alpha \!+\! \omega \alpha^4}} \; , \\ 
f_2\Bigl(X(\zeta)\Bigr) 
& \!=\! & -6 \omega \int_0^{\alpha} \!\!\! d\alphaa \, 
\frac1{\alphaa^{2} \sqrt{1 \!+\! \alphaa \!+\! \omega \alphaa^{4}} 
\, \overline{I}(\alphaa)} \int_0^{\alphaa} \!\!\! d\alphaaa \, 
\frac{\alphaaa^{ 4} \, \overline{I}(\alphaaa)}{\sqrt{1 \!+\! 
\alphaaa \!+\! \omega \alphaaa^{ 4}}} \; ,\qquad \\
f_3\Bigl(X(\zeta)\Bigr) 
& = & -12 \omega \int_0^{\alpha} \!\! d\alphaa \, \frac1{\alphaa^{2} 
\sqrt{1 \!+\! \alphaa \!+\! \omega \alphaa^{4}} \, \overline{I}(\alphaa)} \\
& & \hspace{2cm} \times
\int_0^{\alphaa} \!\! d\alphaaa \, \frac{(\alphaa \!+\! \omega \alphaa^{4}
\!-\! \alphaaa \!-\! \omega \alphaaa^{4} ) \, 
\overline{J}(\alphaaa)}{\sqrt{1 \!+\! \alphaaa \!+\! \omega 
\alphaaa^{4}}} \; , \qquad \label{f32} 
\end{eqnarray}
and one can use these expressions to numerically evolve 
$f\left(X(\zeta)\right)$. 

Having thus obtained $f\left(X(\zeta)\right)$ as a function of $\zeta$, 
the last step is to get $\zeta$ as function of $X$. Before doing so, it 
is also of interest to check that nothing goes wrong at late times. For 
that purpose we need the following small $\zeta$ expansions,
\begin{eqnarray}
h(\zeta) & = & \sqrt{\Omega_{\Lambda}} + \frac{\Omega_{m}}{2 
\sqrt{\Omega_{\Lambda}}} \, \zeta^3 + O(\zeta^4) \; , \\ 
I(\zeta) & = & \frac{4 \sqrt{\Omega_{\Lambda}}}{ \zeta^3} -
\frac{3 \Omega_{m}}{\sqrt{\Omega_{\Lambda}}} \, \ln(\zeta) + O(1) \; , \\ 
\Phi(\zeta) & = & -\frac1{\zeta^2} + O(1) \; .
\end{eqnarray}
Now substitute these into each of the three terms in expression (\ref{f2prime})
for $d f/d\zeta$ to get 
\begin{eqnarray}
\frac{d f}{d\zeta} = -\frac2{\zeta} + O(\zeta) + \Biggl[\frac{\zeta^5}{
4 \Omega} + O\Bigl(\zeta^8 \ln(\zeta)\Bigr)\Biggr] \Biggl\{\frac{8 
\Omega_{\Lambda}}{\zeta^6} + O\Bigl(\frac{\ln(\zeta)}{\zeta^3}\Bigr)
\Biggr\} = O(\zeta) \; .
\end{eqnarray}
This implies that $f$ approaches a constant at late times.  
In the above equation (\ref{ffin}),  the three contributions appearing in the right hand side  can be expressed using elliptic integrals.  

\begin{figure}
\begin{center}
\includegraphics{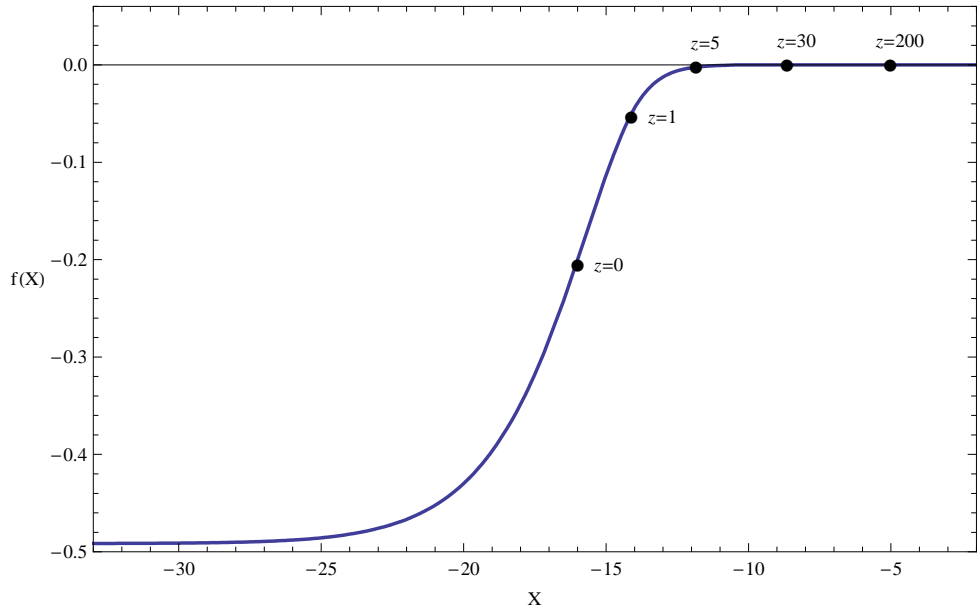}
\caption{Plot (solid blue curve) of the reconstructed function $f(X)$ for 
the non local cosmology reproducing $\Lambda$CDM background cosmological 
evolution, with the same matter content but no cosmological constant. The 
parameters corresponding to the background cosmology are those of the 
latest WMAP release \cite{WMAP}. Circles indicate values of the function 
$f(X)$ with the corresponding value of the redshift $z$ indicated above.}
\label{Fig1}
\end{center}
\end{figure}

Let us now turn to obtaining an expression for $X(\zeta)$. This also 
involves an elliptic integral. Indeed, equation (\ref{DEFX}) reads 
\ba \label{Xzeta}
X = - \int^{\infty}_\zeta \frac{d \zetaa \zetaa^{2}}{H(\zetaa)} 
\int^\infty_{\zetaa} \frac{d \zetaaa}{\zetaaa^{4} H(\zetaaa)} R(\zetaaa)
\equiv - \int^{\infty}_\zeta \frac{d \zetaa\zetaa^{2}}{h (\zetaa)} 
I\left(\zetaa\right),
\ea
where $I$ is defined as in equation (\ref{Idef}). For a 
chosen set of parameters $\{\Omega_\Lambda, \Omega_m, \Omega_r\}$,  
numerical evaluations of the right hand sides of equation (\ref{Xzeta}) 
and (\ref{ffin}) can easily be obtained, from which one can get $f$ as a 
function of $X$. The result is plotted in figure \ref{Fig1} for  
$\{\Omega_\Lambda, \Omega_m, \Omega_r\}$ = $\{0.72,0.28,8.5\times 
10^{-5}\}$ which correspond to the latest WMAP values \cite{WMAP}.

A simple analytic parameterization $f_{\rm an}$ of the found function 
$f$ is given by
\ba
f_{\rm an}(X) = 0.245 \left[
   \tanh(0.350 Y + 0.032 Y^2 + 0.003 Y^3)-1\right], \label{analytic}
\ea
where $Y$ is defined by $Y \equiv (X+16.5)$. It would hardly be 
distinguishable from the numerical solution should it be plotted 
together with the latter on figure \ref{Fig1}. These numerical and
analytic expressions for $f(X)$ allow us to resolve one of the major
open questions about this class of models: do they make significant
corrections to general relativity when expanded around flat space?
The answer is ``no.'' One can see from Figure~1 that the curve is
almost flat near $X = 0$. From the analytic expression (\ref{analytic})
we compute,
\begin{eqnarray}
f_{\rm an}'(0) & = & .245 \Bigl[0.350 + 0.064 Y_0 + 0.009 Y_0^2\Bigr]
\nonumber \\
& & \hspace{3cm} \times
{\rm sech}^2\Bigl[0.350 Y_0 + 0.032 Y_0^2 + 0.003 Y_0^3\Bigr] \; , \\
& \sim & 10^{-24} \; ,
\end{eqnarray}
where $Y_0 = 16.5$. So we find an utterly negligible linear correction.

\section{Discussion}

In this work we have presented a general method to reproduce a given 
arbitrary cosmological evolution from a distorted non local form of the 
action for gravity as presented in Ref. \cite{DW}. This method was 
applied here to $\Lambda$CDM cosmology and we obtained the distortion 
function $f$ that leads, via action (\ref{DL2}), to exactly the same 
cosmological evolution as in $\Lambda$CDM, with the same matter content 
but no cosmological constant. It is very interesting to note that the 
function we obtain numerically is almost indistinguishable from a simple
analytic form (\ref{analytic}). To be sure, this function contains some 
free parameters --- for example, the value $X = -16.5$ where the $\tanh$
passes through zero, or the scaling of its full variation by $0.49$. 
However, these are all dimensionless numbers of order one. This is a 
consequence of two crucial properties of nonlocal models of type (\ref{FGR}):
\begin{itemize}
\item{The onset of late time acceleration is triggered by the very
recent cosmological transition from $R \approx 0$ during radiation domination 
to $R \sim 1/t^2$ during matter domination; and}
\item{Even after this transition the nonlocal operator $\frac{1}{\square} R
\sim -\ln(t/t_{\rm eq})$ grows very slowly.}
\end{itemize}

Several very interesting questions are left for future work. First, as 
far as cosmology is concerned, a natural question to address is if the 
model which gives the same background evolution as $\Lambda$CDM can be 
distinguished from $\Lambda$CDM by considering observables that contain 
information in addition to the background evolution. This requires in 
particular working out the theory of cosmological perturbations for the 
non local model (see \cite{TK2}). To do so, a good starting point is in 
fact this work, using for example, the analytic form (when necessary) of 
the reconstructed distortion function. Other issues concern the various 
tests of gravity one can consider, in particular those in the solar 
system or those involving binary pulsars. It would be extremely 
interesting to apply those tests to the framework of Ref. \cite{DW}, and 
a first investigation along these lines has already been carried out in 
Ref.  \cite{TK2}. Note in particular that the way cosmic acceleration is 
produced in the model considered here, is via a strengthening of the 
Newton constant encoded in the non local function $f$.  However, the 
strengthening of the Newton constant we mentioned, strictly speaking only 
applies to cosmological distances, and things would be radically different, 
hence requiring a completely different analysis, inside gravitationally 
bound objects such as a galaxy or a cluster. This raises various questions 
about the effects of the non local operator $\square^{-1}$ inside matter.

\label{sec5}
\section*{\bf Acknowledgements}
RPW is grateful for the hospitality of the Laboratory APC, CNRS,
University Paris 7, where a major part of this project was done.
This work was partially supported by European Union grant 
INTERREG-IIIA, by NSF grant PHY-0653085, by the Institute for 
Fundamental Theory at the University of Florida, as well as the JCJC 
ANR grant "MODGRAV".

\end{document}